\documentclass[conference]{IEEEtran}
\normalsize
\ifCLASSINFOpdf
\else
\fi

\usepackage{amsthm}
\usepackage{amsmath}
\usepackage{amssymb}
\usepackage{graphicx}
\usepackage{esint}
\usepackage{amsfonts}
\usepackage{cite}
\usepackage{balance}
\usepackage{caption}
\usepackage{subcaption}
\usepackage{epstopdf}
\usepackage{color}
\usepackage{tikz}

\def\BibTeX{{\rm B\kern-.05em{\sc i\kern-.025em b}\kern-.08em
		T\kern-.1667em\lower.7ex\hbox{E}\kern-.125emX}}

\makeatletter

\theoremstyle{plain}
\newtheorem{prop}{\protect\theoremname}

\makeatother

\providecommand{\theoremname}{Proposition}

\theoremstyle{plain}

\makeatother

\providecommand{\lemmaname}{Lemma}
\DeclareMathOperator{\erf}{erf}

\begin{document}
%\markboth{Submitted to \textit{IEEE Communications Letters}}{Boulogeorgos, Chatzidiamantis, Karagiannidis, Spectrum Sensing in Multiple Primary Users Environment under Nakagami-m Fading}
\title{ Performance Analysis of Multi-Reconfigurable Intelligent Surface-Empowered THz Wireless~Systems }

\author{%\vspace{-0.2cm} 
Alexandros-Apostolos A. Boulogeorgos$\,^{1}$, Nestor Chatzidiamantis$\,^{2}$,  Harilaos G. Sandalidis$\,^{3}$, Angeliki Alexiou$\,^{1}$, \\ and Marco Di Renzo$\,^{4}$  %and Leonidas Georgiadis, \IEEEmembership{Senior, IEEE,}   
%\thanks{This work was presented in part at IEEE ICC 2015 - Seventh Workshop on Cooperative and Cognitive Networks (CoCoNet7). }
\\
\footnotesize{$\,^{1}$Digital Systems, University of Piraeus Piraeus 18534 Greece.} e-mails: al.boulogeorgos@ieee.org, alexiou@unipi.gr.
\\
$\,^{2}$Department of Electrical and Computer Engineering, Aristotle University of Thessaloniki, 54124 Thessaloniki, Greece. e-mail: nestoras@auth.gr.
\\
$\,^{3}$Department of Computer Science and Biomedical Informatics, University of Thessaly, 35131 Lamia, Greece, 
e-mail: sandalidis@dib.uth.gr.\\
$\,^{4}$Universit\'e Paris-Saclay, CNRS, CentraleSup\'elec, Laboratoire des Signaux et Syst\`emes, 3 Rue Joliot-Curie, 91192 Gif-sur-Yvette, France. \\
e-mail: marco.di-renzo@universite-paris-saclay.fr

}
\maketitle	
%\vspace{-0.4cm}
\begin{abstract}
 In this paper, we introduce a theoretical framework for analyzing the performance of multi-reconfigurable intelligence surface (RIS) empowered terahertz (THz) wireless systems subject to turbulence and stochastic beam misalignment. In more detail, we extract a closed-form expression for the outage probability that quantifies the joint impact of turbulence and misalignment as well as the effect of transceivers' hardware imperfections. Our results highlight the importance of accurately modeling both turbulence and misalignment when assessing the performance of multi-RIS-empowered THz wireless~systems.        
\end{abstract}
%\vspace{-0.2cm}
\begin{IEEEkeywords}
Outage probability, performance analysis, reconfigurable intelligent surfaces, THz wireless communications.
\end{IEEEkeywords}

%\setlength{\abovedisplayskip}{1.1pt} \setlength{\abovedisplayshortskip}{1.2pt}
%\setlength{\belowdisplayskip}{1.1pt} \setlength{\belowdisplayshortskip}{1.2pt}

%\vspace{-0.5cm}
\section{Introduction}\label{S:Intro}
%\vspace{-0.2cm}

The sixth-generation wireless era (6G) is envisioned to bring the fiber quality-of-experience into the wireless world, by supporting aggregated data-rates that may exceed $1\text{ }\mathrm{Tb/s}$ with outage probability in the order of $10^{-6}$~\cite{Boulogeorgos2018,WP:Wireless_Thz_system_architecture_for_networks_beyond_5G,Boulogeorgos2021a}. To achieve this challenging goal, both the research and industrial societies turned their attention into the yet-unregulated terahertz (THz) band~\cite{Zhang2019,Tsiftsis2022,C:UserAssociationInUltraDenseTHzNetworks}. On the one hand, THz links can provide the required bandwidth, however, they come with an inherent disadvantage; they are sensitive to blockage~\cite{Zhang2018,Boulogeorgos2022book,Stratidakis2020}. To counterbalance blockage, the idea of creating alternative paths between the source (S) and final destination (D) through reconfigurable intelligent surfaces (RISs) has been considered~\cite{Renzo2020,Tsilipakos2020,Boulogeorgos2020a,Boulogeorgos2022a}.

Inspired by this, a great amount of effort has been put on designing RIS that operate in the THz band~\cite{Cai2018,Venkatesh2020,Amin2021,Ojaroudi2021} as well as theoretically characterizing the corresponding system performance~\cite{Tekbiyik2020,Du2020,Boulogeorgos2021}. In particular, in~\cite{Cai2018}, a vanadium dioxide ($\text{VO}_2$)-based multi-functional RIS was presented, while, in~\cite{Venkatesh2020}, a large-scale RIS that employs arrays of complementary metal-oxide-semiconductor (CMOS)-based chip tiles operating at $0.3\text{ }\mathrm{THz}$ was reported. In~\cite{Amin2021}, a graphene plasmonic metasurface was documented. On~\cite{Tal2020}, a broadband plasmonic metasurface emitter, operating at the THz band,  was presented. The design of a graphene-based RIS structure with beam steering and focusing capabilities at $4.35\text{ }\mathrm{THz}$ was discussed in~\cite{Ojaroudi2021}. From the performance analysis points of view, the error performance of RIS-empowered THz wireless satellite systems in the presence of antenna misalignment was studied in~\cite{Tekbiyik2020}. In~\cite{Du2020}, the authors assessed the joint impact of hardware imperfections and antenna misalignment on RIS-empowered indoor THz wireless systems. The coverage performance of RIS-empowered THz wireless systems was quantified in~\cite{Boulogeorgos2021}.

Recently, the idea of using multi-RISs to ensure uninterrupted connectivity between the transmitter and the receiver was reported in~\cite{Liaskos2018}. Despite the  features and additional degrees of freedom that this idea can enable, to the best of the authors knowledge, a theoretical framework for analyzing the achievable performance of multi-RIS THz wireless systems has not been formulated yet. Motivated by this,  we present an analytical framework, which covers the aforementioned gap by extracting closed-form expressions for the outage probability (OP) of such systems in the presence of turbulence and beam misalignment. This framework also accounts the impact of transceiver's hardware imperfections.

\subsubsection*{Notations} 
The product of $x_1\,x_2\,\cdots\, x_L$ is $\prod_{l=1}^{L}x_l$. 
 $\Pr\left(\mathcal{A}\right)$ stands for the probability for the event $\mathcal{A}$ to be valid. 
The modified Bessel function of the second kind of order $n$ is represented as~$\mathrm{K}_n(\cdot)$~\cite[eq. (8.407/1)]{B:Gra_Ryz_Book}. 
The  Gamma~\cite[eq. (8.310)]{B:Gra_Ryz_Book} function is  denoted by  $\Gamma\left(\cdot\right)$, and the error-function is represented by $\erf\left(\cdot\right)$~\cite[eq. (8.250/1)]{B:Gra_Ryz_Book}. Finally,  $G_{p, q}^{m, n}\left(x\left| \begin{array}{c} a_1, a_2, \cdots, a_{p} \\ b_{1}, b_2, \cdots, b_q\end{array}\right.\right)$  returns the Meijer G-function~\cite[eq. (9.301)]{B:Gra_Ryz_Book}.

\section{System model}

\begin{figure}
	\centering
	\vspace{-0.3cm}
	\scalebox{.55}{\input{systemModel}}
	\vspace{-0.25cm}
	\caption{Cascaded multi-RIS empowered THz wireless model.}
	\vspace{-0.8cm}
	\label{Fig:SM}
\end{figure} 

As illustrated in~Fig.~\ref{Fig:SM}, we consider a multi-RIS-empowered outdoor THz wireless systems that consists of a single $S$ and $D$ as well as $N$ RISs. Both $S$ and $D$ are equipped with high-directional antennas. We assume that the direct link between the $S$ and $D$ is blocked. Each RIS acts as a reflector that steers the incident beam towards the desired direction. The baseband equivalent received signal at $D$ can be expressed~as~\cite{PhD:Boulogeorgos}
\begin{align}
	r = A \left(x + \eta_{s}\right) + \eta_{d} + n.
	\label{Eq:y}
\end{align}    
In~\eqref{Eq:y}, $x$ is the transmission signal, $n$ stands for the additive white Gaussian noise (AWGN) and is modeled as a zero-mean complex Gaussian (ZMCG) process with variance $N_o$. The $S$ and $D$ distortion noises, due to transmitter's and receiver's hardware imperfections, are respectively modeled by two independent RVs that are respectively denoted by $\eta_{s}$ and $\eta_{d}$. According to~\cite{Boulogeorgos2020b,Boulogeorgos2020,B:Schenk-book}, for a given channel realization,  $\eta_{s}$ and $\eta_{d}$ are ZMCG processes with variances~\cite{A:LC_CR_vs_SS,Boulogeorgos2019,C:Energy_Detection_under_RF_impairments_for_CR}
\begin{align} 
	\sigma_{s}^2=\kappa_{s}^2 P_s \text{ and } \sigma_{d}^2=\kappa_{r}^2 A^2 P_s,
	\label{Eq:sigma_of_distortion}
\end{align} 
 where $\kappa_{t}$ and $\kappa_{r}$ denote the error vector magnitude of the $S$'s transmitter and the $D$'s receiver, while $P_s$ represents the transmission power. 

Likewise, in~\eqref{Eq:y},
\begin{align}
	A = \prod_{i=1}^{L} h_{p,i} h_{t,i} g_i 
	\prod_{i=L+1}^{N} h_{t,i} g_i. 
	\label{Eq:A_t}
\end{align}  
stands for the end-to-end (e2e) channel coefficient, while $h_{p,i}$ and $h_{t,i}$ are  the misalignment fading and the Gamma-Gamma (GG) distributed turbulence coefficients that can be described from the following probability density functions (PDFs):
\begin{align}
	f_{h_{p,i}}(x) = \frac{\xi_i}{A_{o,i}^{\xi_i}} x^{\xi_i-1},\text{ with } 0\leq x \leq A_{o,i},
	\label{Eq:f_l_i}
\end{align} 
and
\begin{align}
	f_{h_{t,i}}(x) &= \frac{2}{\Gamma\left(\alpha_i\right) \Gamma\left(\beta_i\right)} \left(\frac{\alpha_i \beta_i}{\Omega_i}\right)^{\frac{\alpha_i + \beta_i}{2}}  x^{\frac{\alpha_i + \beta_i}{2}-1}
	\nonumber \\
	\times & \mathrm{K}_{\alpha_i-\beta_i}\left(\alpha_i-\beta_i, 2 \sqrt{\frac{\alpha_i \beta_i x}{\Omega_i}} \right),
	\label{Eq:f_r_i}
\end{align} 
respectively. In~\eqref{Eq:f_l_i}, 
\begin{align}
	A_{o,i} &= [\erf\left(\upsilon_i\right)]^2
	\label{Eq:A_0}
\end{align}
represents the fraction of the power captured by the $i-$th RIS or the $D$ in the ideal case of zero radial displacement. 
Moreover,    
\begin{align}
	\upsilon_i = \frac{\sqrt{\pi}b_i}{\sqrt{2}w_{d_i}}.
	\label{Eq:v}
\end{align}
with $b_i$ and $w_{d_i}$  representing the radius of the circular aperture at the $i-$th RIS, with $i\in[1, N-1]$, or $D$'s side, for $i=N$, respectively and the beam waste  on the corresponding plane. 
Additionally,  $\xi_i$ is the equivalent beam radius, $w_{\mathrm{eq},i}$, to the pointing error displacement standard deviation square ratio and can be evaluated~as 
\begin{align}
	\xi_i=\frac{w_{\mathrm{eq},i}^2}{4\sigma_{s,i}^2}
	\label{Eq:xi}
\end{align}
with $\sigma_{s,i}^2$ denoting  the pointing error displacement (jitter) variance. 

The $\alpha_i$ and $\beta_i$ parameters of~\eqref{Eq:f_r_i} can be computed according to~\cite{Taherkhani2020}~as
\begin{align}
	\alpha_i &= {\left(\exp\left(\frac{0.49\sigma_{R_i}^2}{\left(1+0.65 D_i^2+1.11\sigma_{R_i}^{12/5}\right)^{7/6}} \right)-1 \right)^{-1}} 
\end{align} 
%and
\begin{align}
	\beta_i &= \left(\exp\left( \frac{0.51\sigma_{R_i}^2\left(1+0.69\sigma_{R_i}^{12/5}\right)^{-5/6}}{1+0.9 D_i^2 +0.62 D_i^2 \sigma_{R_i}^{12/5}} \right)-1 \right)^{-1}
\end{align}
while
\begin{align}
	\sigma_{R_i}^2=1.23 C_n^2 \left(\frac{2\pi}{\lambda}\right)^{7/6} d_i^{11/6},
\end{align} 
with $\lambda$ being the wavelength $C_n^2$ the reflection index structure parameter, which, according to ITU-T can be evaluated as in~\cite{Hill1988}. Additionally, 
\begin{align}
	D_i = \sqrt{\frac{\pi b_i^2}{2\lambda d_i}}.
\end{align}
 Finally, $g_i$ represents the deterministic path-gain coefficient of the $i-$th link and can be written~as
\begin{align}
	g_i =  g_{f,i}\left(f, d_i\right) \tau\left(f, d_i\right),
\end{align} 
where 
\begin{align}
	g_{f,i}\left(f, d_i\right)=\left\{ \begin{array}{l l}
		\frac{c}{4\pi f d_i} \sqrt{G_{s}}, & \text{ for } i=1 \\
		\frac{c}{4\pi f d_i} R_{i-1}, & \text{ for } i\in[2, N-1] \\
		\frac{c}{4\pi f d_i} \sqrt{G_{d}}, & \text{ for } i=N 
	\end{array}\right.,
	\label{Eq:g_f}
\end{align}
is the free space path loss coefficient,
and 
\begin{align}
	\tau\left(f, d_i\right) = \exp\left(-\frac{1}{2}\kappa(f)\, d_i\right).
	\label{Eq:tau}
\end{align}
stands for the molecular absorption gain coefficient.
In~\eqref{Eq:g_f}, $f$ and $c$  are the transmission frequency and the speed of light, respectively; $G_{s}$, $R_{i}$ and $G_{d}$ represent the $S$ transmission antenna gain, the $i-$th RIS reflection coefficient, and the $D$ reception antenna~gain, respectively. Likewise, $\kappa(f)$ is the molecular absorption coefficient, which depends on the atmospheric temperature, pressure, as well as relative humidity and can be evaluated as in~\cite[eq. (8)]{A:Analytical_Performance_Assessment_of_THz_Wireless_Systems}.

\section{Outage probability}
From~\eqref{Eq:y}, the signal-to-distortion-plus-noise-ratio (SDNR) can be  obtained~as 
\begin{align}
	\gamma = \frac{A^2 P_s}{A_{t}^2 \left(\kappa_t^2 + \kappa_r^2 \right) P_s + N_o},
	\label{Eq:gamma}
\end{align}
where $N_o$ is the noise variance. 

The following proposition returns the outage probability of the multi-RIS-empowered THz wireless system.
\begin{prop}
	The outage probability of the multi-RIS-empowered THz wireless system can be evaluated as in~\eqref{Eq:P_o_THz_s3}, given at the top of the next page.
\begin{figure*}
 \begin{align}
 	P_o = \left\{\begin{array}{l l}	
 	\frac{\prod_{i=1}^{L}\xi_i}{\prod_{i=1}^N\Gamma\left(\alpha_i\right) \Gamma\left(\beta_i\right)} & \\ 
 	\times\mathrm{G}_{L+1, 2N+L+1}^{2N+L, 1}\left(\frac{1}{\prod_{i=1}^{N}\frac{\Omega_i}{\alpha_i\beta_i}\prod_{i=1}^{L}A_{o,i}} \frac{1}{\prod_{i=1}^{N}g_i} \frac{1}{\sqrt{1-\gamma_{\mathrm{th}}\left(\kappa_t^2 + \kappa_r^2\right)}} \sqrt{\frac{\gamma_{\mathrm{th}}}{\gamma_{s}}} 
 	\left|\begin{array}{c}1, \xi_1+1, \cdots, \xi_L+1 \\  \alpha_1, \cdots, \alpha_N, \beta_1, \cdots, \beta_N, \xi_1, \cdots, \xi_L, 0  \end{array}\right.\right),
 	& \\
 	& \hspace{-5cm} \text{ for } \gamma_{\mathrm{th}} \leq \frac{1}{\kappa_{t}^2 + \kappa_r^2}\\
 	1, & \hspace{-5cm} \text{ for } \gamma_{\mathrm{th}} > \frac{1}{\kappa_{t}^2 + \kappa_r^2} 
 	\end{array}
 	\right.
 	\label{Eq:P_o_THz_s3} 
 \end{align}
\hrulefill
\end{figure*} 
In~\eqref{Eq:P_o_THz_s3}, $\gamma_s$ is the transmission SNR multiplied by the deterministic path-gain, i.e., 
\begin{align}
	\gamma_s = \frac{P_s \prod_{i=1}^{N} g_i}{N_o}.
\end{align}
\end{prop}
\begin{IEEEproof}
	For brevity, the proof of the proposition is given in the Appendix. 
\end{IEEEproof}
Note that from~\eqref{Eq:P_o_THz_s3}, it becomes evident that hardware imperfections set a limit to the maximum allowed spectral efficiency of the transmission scheme. In more detail, since the spectral efficiency, $p$, is connected with the SNR threshold through 
\begin{align}
p=\log_{2}\left(1+\gamma_{th}\right),
\end{align}
choosing a $p$  greater than $\log_{2}\left(1+\frac{1}{\kappa_t^2+\kappa_r^2}\right)$ would lead to an OP equal to~$1$. 

 \section{Results \& Discussion}\label{S:Results}
 
 This section verifies the presented theoretical framework by means of  Monte Carlo simulations. In what follows, lines are used to denote analytical results, while markers are employed for simulations. Unless otherwise stated, the following insightful scenario is considered.  The relative humidity, temperature, and pressure  respectively are set to $50\%$, $273\,^{o}K$, and $101325\text{ }\mathrm{Pa}$.  Also, $G_s=G_t=50\text{ }\mathrm{dBi}$, $R_i=1$ for $i\in[1, N-1]$ and~$f=300\text{ }\mathrm{GHz}$. The transmission power, bandwidth, $B$, and spectral efficiency of the transmission scheme are respectively set to $0\text{ }\mathrm{dBW}$, $50\text{ }\mathrm{GHz}$ and $2\text{ }\mathrm{bit/s/Hz}$. Moreover, $d_1=d_2=100\text{ }\mathrm{m}$. At the destination side, we assumed that the low noise amplifier's gain and noise figure (NF) are $35\text{ }\mathrm{dB}$, and $1\text{ }\mathrm{dB}$, respectively. The mixer and miscellaneous losses are respectively $5$ and $3\text{ }\mathrm{dB}$. The mixer's NF is $6\text{ }\mathrm{dB}$. The thermal noise power is evaluated as $N_1 = k_B\, T\,B$, where $k_B$ is the Boltzman's constant.

 \begin{figure}
	\centering\includegraphics[width=0.9\linewidth,trim=0 0 0 0,clip=false]{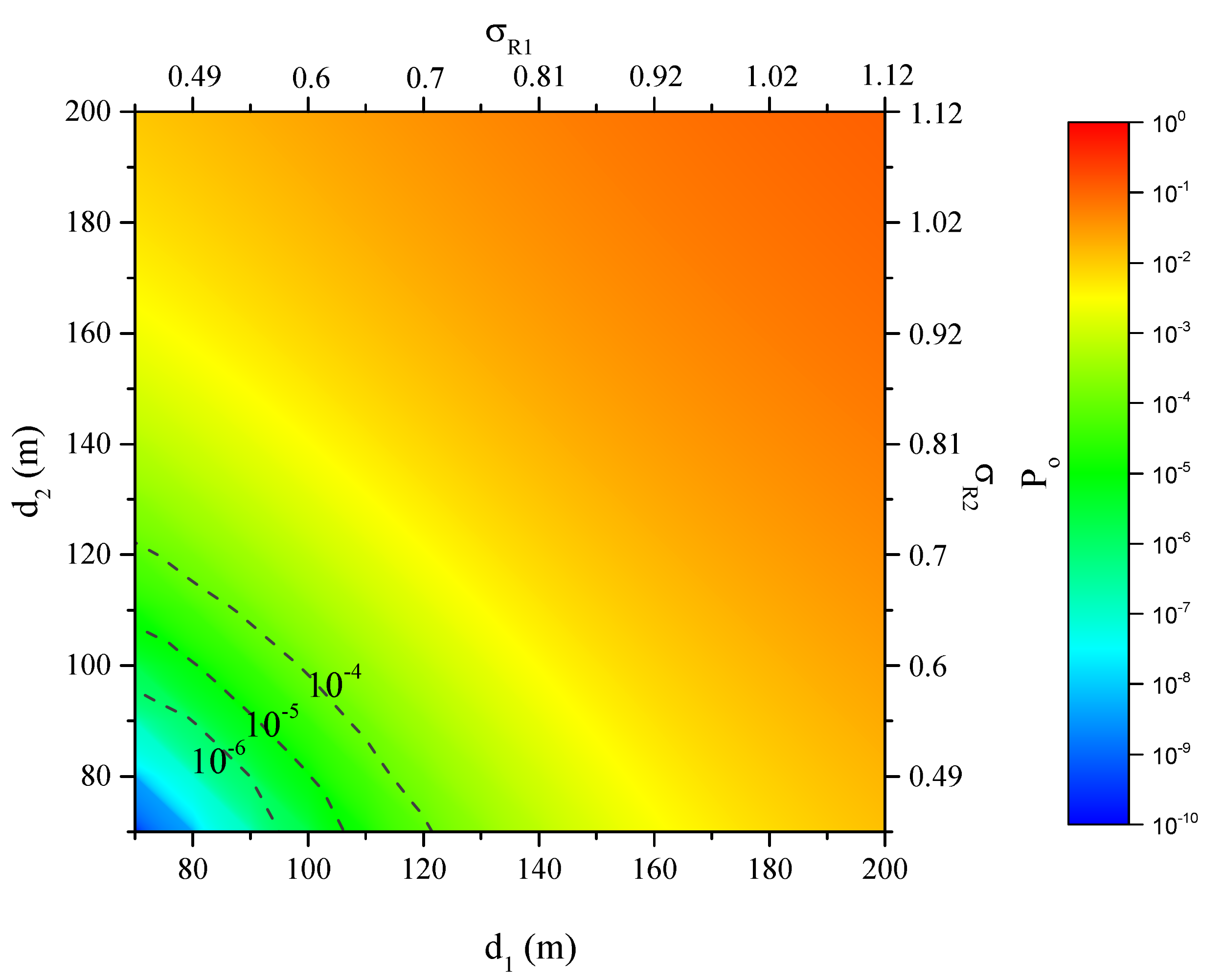}\vspace{-0.2cm}
	\caption{OP vs $d_1$ and $d_2$.}
	\label{Fig:OP_d1_d2}
\end{figure}
Figure~\ref{Fig:OP_d1_d2} quantifies the impact of turbulence on the outage performance of multi-RIS-empowered THz wireless systems in the absence of misalignment and hardware imperfections, assuming $N=2$. In more detail, the OP is illustrated as a function of $d_1$ and $d_2$. In this figure, for the sake of convenience, the corresponding values of $\sigma_{R1}$ and $\sigma_{R2}$ are respectively provided in the top horizontal and right vertical axes.   As expected, for a fixed $d_1$, as $d_2$ increases, $\sigma_{R2}$ increases; thus, an outage performance degradation is observed.  Similarly, for a given $d_2$, as $d_1$ increases, $\sigma_{R1}$ also increases, i.e., turbulence intensity increases; in turn, the OP increases.  Finally, from this figure, we observe that for a given e2e transmission distance, $d_1+d_2$, the worst outage performance is observed for $d_1=d_2$. For example, for $d_1+d_2=300\text{ }\mathrm{m}$, the OP for the case in which $d_1=100\text{ }\mathrm{m}$ and $d_2=200\text{ }\mathrm{m}$ is equal to $2.6\times 10^{-2}$, while, for $d_1=d_2=150\text{ }\mathrm{m}$, it is~$2.7\times 10^{-2}$.

\begin{figure}
	\centering\includegraphics[width=0.8\linewidth,trim=0 0 0 0,clip=false]{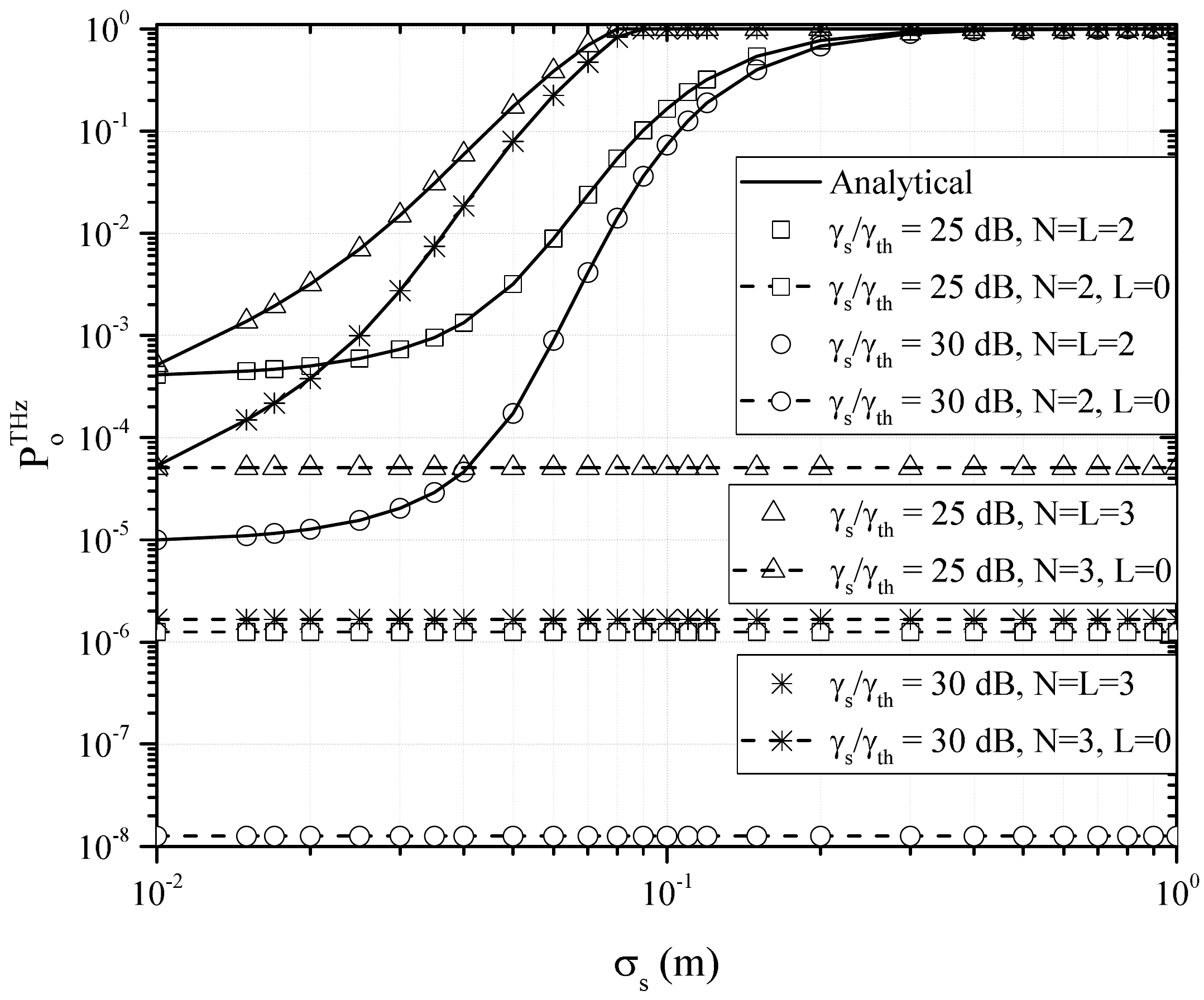}\vspace{-0.2cm}
	\caption{OP vs $\sigma_s$, for different values of $N$, $L$ and $\gamma_s/\gamma_{\mathrm{th}}$.}
	\label{Fig:OP_N_L_gs}
\end{figure}

 Figure~\ref{Fig:OP_N_L_gs} illustrates the OP as a function of $\sigma_s$, for different values of $N=L$ and $\gamma_{s}/\gamma_{\mathrm{th}}$ in the presence of turbulence, assuming that $\kappa_t=\kappa_r=0$, and the transmission distances for all the links are set to $100\text{ }\mathrm{m}$. In this scenario, we assume that $\sigma_{s,1}=\sigma_{s,2}=\cdots=\sigma_{s,L}=\sigma_s$. As a benchmark, the OP in the absence of misalignment fading is also plotted. As expected, for given $N=L$ and $\gamma_{s}/\gamma_{\mathrm{th}}$, the outage performance degrades, as the intensity of misalignment fading, i.e., $\sigma_s$, becomes more severe.  Likewise, for fixed $N=L$ and $\sigma_s$, as $\gamma_{s}/\gamma_{\mathrm{th}}$, the OP decreases. On the other hand, for given $\sigma_s$ and $\gamma_{s}/\gamma_{\mathrm{th}}$, as $N=L$ increases, the OP also increases. For instance, for $\sigma_s=4\text{ }\mathrm{cm}$ and $\gamma_{s}/\gamma_{\mathrm{th}}=25\text{ }\mathrm{dB}$, the OP increases by approximately $10$ times, as $N=L$ changes from $2$ to $3$. Finally, from this figure, the detrimental impact of misalignment fading becomes apparent by comparing; thus, the importance of accurately characterizing the channels of the RIS-empowered THz wireless systems is highlighted. 
 
 \begin{figure}
 	\centering\includegraphics[width=0.9\linewidth,trim=0 0 0 0,clip=false]{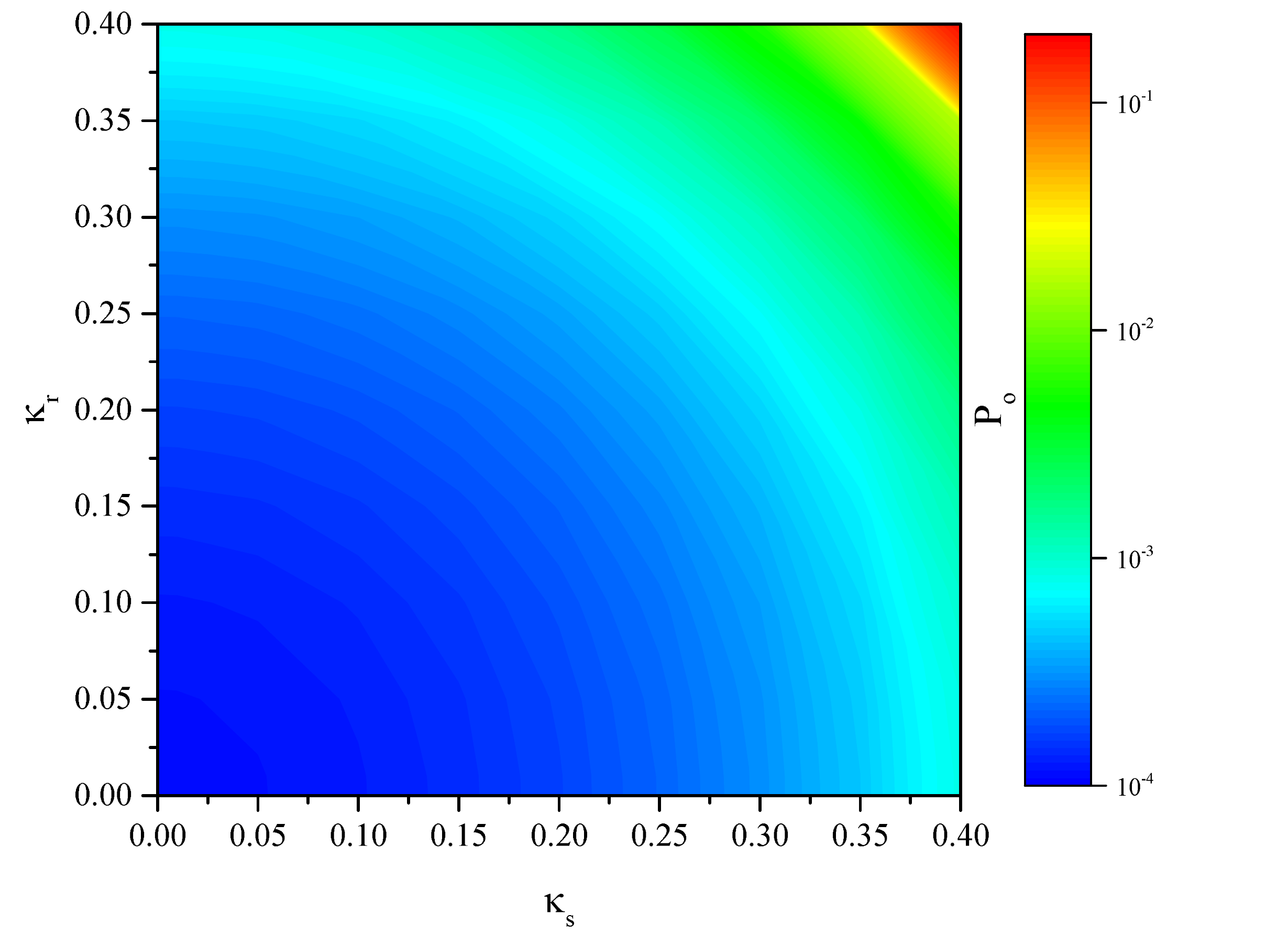}\vspace{-0.2cm}
 	\caption{OP vs $\kappa_t$ and $\kappa_r$.}
 	\label{Fig:OP_HW}
 \end{figure}

Figure~\ref{Fig:OP_HW} presents the impact of hardware imperfections in multi-RIS-empowered THz wireless systems, assuming $N=2$, and $L=0$. Note that the case in which $\kappa_t=\kappa_r=0$ corresponds to the case in which both the $S$ and $D$ are equipped with ideal RF front-ends. From this figure, it becomes apparent that, for a given $\kappa_{t}$, as $\kappa_{r}$ increases, the OP also increases. Similarly, for a fixed $\kappa_{t}$, an outage performance degradation occurs, when $\kappa_{r}$ increases. Additionally, for a constant $\kappa_{t}+\kappa_r$, we observe that the OP is maximized for $\kappa_{t}=\kappa_{r}$. Finally,  it is verified that systems with the same $\kappa_{t}^2+\kappa_r^2$ achieve the same~OP.

 \section{Conclusions}\label{S:Conclusions}
 In this paper, we presented a theoretical framework for the characterization of the outage performance of multi-RIS-empowered THz wireless systems in the presence of turbulence and misalignment that accounts the impact of transceivers hardware imperfections. In this direction, a novel closed-form expressions for the system's OP was extracted. Our results revealed that as the number of RIS and thus the distance between S and D increases, the turbulence and misalignment fading intensity becomes more severe; as a result, the outage performance degrades. Likewise, the importance of considering the impact of unavoidable transceivers' hardware imperfections was~emphasized.

  \section*{Appendix}
  \section*{Proof of Proposition 1}
  
  The OP is defined~as
  \begin{align}
  	P_o = \Pr\left(\gamma\leq \gamma_{\mathrm{th}}\right),
  	\label{Eq:P_o_THz}
  \end{align} 
  where $\gamma_{\mathrm{th}}$ represents the SNR threshold. By applying~\eqref{Eq:gamma} in~\eqref{Eq:P_o_THz}, we get 
  \begin{align}
  	P_o = \left\{\begin{array}{l l} \Pr\left( A^2 \leq \frac{1}{1-\gamma_{\mathrm{th}}\left(\kappa_t^2 + \kappa_r^2\right)} \frac{\gamma_{\mathrm{th}}}{\gamma_{s}} \right),  & \text{ for } \gamma_{\mathrm{th}} \leq \frac{1}{\kappa_{t}^2 + \kappa_r^2} \\ 1, & \text{ for } \gamma_{\mathrm{th}} > \frac{1}{\kappa_{t}^2 + \kappa_r^2} \end{array}\right.,
  	\label{Eq:P_o_THz_s2} 
  \end{align}
or equivalent 
\begin{align}
	P_o = 
	\left\{\begin{array}{l l}
		F_{A} \left( 
		\frac{1}{\sqrt{1-\gamma_{\mathrm{th}}\left(\kappa_t^2 + \kappa_r^2\right)}} \sqrt{\frac{\gamma_{\mathrm{th}}}{\gamma_{s}}} \right),  & \text{ for } \gamma_{\mathrm{th}} \leq \frac{1}{\kappa_{t}^2 + \kappa_r^2}
\\		
1, & \text{ for } \gamma_{\mathrm{th}} > \frac{1}{\kappa_{t}^2 + \kappa_r^2} 
	\end{array}\right.,
\label{Eq:P_o_THz_s4} 
\end{align}
where $F_{A} \left(\cdot \right)$ stands for the cumulative density function (CDF) of~$A$. 

Next, we evaluate the CDF of $A$. In this direction, let us rewrite~\eqref{Eq:A_t}~as
\begin{align}
	A = C\, Y_1 \, Y_2, 
\end{align}
where 
\begin{align}
	C = \prod_{i=1}^{N} g_i,
\end{align}
which is a constant, while
\begin{align}
	Y_1 = \prod_{i=1}^{L} h_{p,i} h_{t,i}
\end{align}
and
\begin{align}
	Y_2 = \prod_{i=L+1}^{N} h_{t,i}. 
\end{align}
In order to derive, a closed-form expression for the CDF of $A_t$, we first need to evaluate the probability density functions (PDFs) of $Y_1$ and $Y_2$. 

For the PDF of $Y_1$, we start the evaluation by assuming that $L=1$. In this case the PDF of $Y_1$ can be obtained~as 
\begin{align}
	f_{Y_1}(x;L=1) = \int_{\frac{x}{A_{o,1}}}^{\infty}\frac{1}{y} f_{h_t}\left(y\right)\, f_{h_p}\left(\frac{x}{y}\right)\,\mathrm{d}x,
\end{align}  
which, by applying~\eqref{Eq:f_r_i} and~\eqref{Eq:f_l_i}, can be rewritten~as
\begin{align}
	f_{Y_1}(x;L=1) &= 2\left(\frac{\alpha_1 \beta_1}{\Omega_1}\right)^{\frac{\alpha_1+\beta_1}{2}} \frac{1}{\Gamma\left(\alpha_1\right) \Gamma\left(\beta_1\right)} \frac{\xi}{A_{o,1}^{\xi_1}} x^{\xi_1-1} 
	\nonumber \\ & \hspace{-1.2cm} \times
	\int_{\frac{x}{A}}^{\infty} y^{\frac{\alpha_1+\beta_1}{2}-1-\xi_1} \, \mathrm{K}_{\alpha_1-\beta_1}\left(2\sqrt{\frac{\alpha_1 \beta_1 y}{\Omega_1}}\right)\,\mathrm{d}y.
	\label{Eq:f_Y_1_L_1}
\end{align}
By employing~\cite[ch. 2.6]{Mathai1973},~\eqref{Eq:f_Y_1_L_1} can be written~as in~\eqref{Eq:f_Y_1_L_1_s2}, given at the top of the next page.
\begin{figure*}
	\begin{align}
		&f_{Y_1}(x;L=1) = \left(\frac{\alpha_1 \beta_1}{\Omega_1}\right)^{\frac{\alpha_1+\beta_1}{2}} \frac{1}{\Gamma\left(\alpha_1\right) \Gamma\left(\beta_1\right)} \frac{\xi}{A_{o,1}^{\xi_1}} x^{\xi_1-1} 
		\int_{\frac{x}{A_{o,1}}}^{\infty} \left( \frac{1}{y}\right)^{-\frac{\alpha_1+\beta_1}{2}+\xi_1+1} \mathrm{G}_{0, 2}^{2, 0}\left( \frac{\alpha_1 \beta_1 y}{\Omega_1}\left| \frac{\alpha_1-\beta_1}{2}, -\frac{\alpha_1-\beta_1}{2}  \right.\right)\,\mathrm{d}y
		\label{Eq:f_Y_1_L_1_s2}
	\end{align}
	\hrulefill
\end{figure*}
Additionally, by applying~\cite[eq. (2.24.5/3)]{B:Prudnikov_v3},~\eqref{Eq:f_Y_1_L_1_s2} can be analytically expressed~as
\begin{align}
	&f_{Y_1}(x;L=1) = \left(\frac{\alpha_1 \beta_1 }{\Omega_1 A_{o,1}}\right)^{\frac{\alpha_1+\beta_1}{2}} \frac{\xi_1}{\Gamma\left(\alpha_1\right) \Gamma\left(\beta_1\right)}  x^{\frac{\alpha_1+\beta_1}{2}-1} 
	\nonumber \\ & \times
	\mathrm{G}_{1, 3}^{3, 0}\left(\frac{\alpha_1\beta_1 x}{A_o \Omega_1}\left| \begin{array}{c} -\frac{\alpha_1+\beta_1}{2}+\xi_1+1 \\ \frac{\alpha_1-\beta_1}{2}, -\frac{\alpha_1-\beta_1}{2}, -\frac{\alpha_1+\beta_1}{2}+\xi_1  \end{array}\right. \right), 
\end{align}
which, with the aid of~\cite{WS:mathematica_function}, can be rewritten~as
\begin{align}
	&f_{Y_1}(x;L=1) = \frac{\xi_1 x^{-1}}{\Gamma\left(\alpha_1\right) \Gamma\left(\beta_1\right)}\mathrm{G}_{1,3}^{3, 0}\left(\frac{\alpha_1\beta_1 x}{A_{o,1} \Omega_1}\left|\begin{array}{c} \xi_1+1 \\ \alpha_1, \beta_1, \xi_1 \end{array}\right. \right).
	\label{Eq:f_Y_1_L_1_final}
\end{align}

For $L=2$, the PDF of $Y_1$ can be obtained~as
\begin{align}
	&f_{Y_1}(x;L=2) = \int_{0}^{\infty} \frac{1}{y} f_{Y_1}(y;L=1) \, f_{Y_1}\left(\frac{x}{y};L=1\right) \, \mathrm{d}y,
\end{align}
which, by applying~\eqref{Eq:f_Y_1_L_1_final}, can be rewritten as in~\eqref{Eq:f_Y_1_L_2_s1}, given at the top of the next page.
\begin{figure*}
	\begin{align}
		&f_{Y_1}(x;L=2) = \frac{\prod_{i=1}^{2}\xi_i}{\prod_{i=1}^2\Gamma\left(\alpha_i\right) \Gamma\left(\beta_i\right)} x^{-1}
		\int_{0}^{\infty} y^{-1} \mathrm{G}_{1,3}^{3, 0}\left(\frac{\alpha_1\beta_1 y}{A_{o,1} \Omega_1}\left|\begin{array}{c} \xi_1+1 \\ \alpha_1, \beta_1, \xi_1 \end{array}\right. \right) \mathrm{G}_{1,3}^{3, 0}\left(\frac{\alpha_2\beta_2 x}{A_{o,2} \Omega_2 y}\left|\begin{array}{c} \xi_2+1 \\ \alpha_2, \beta_2, \xi_2 \end{array}\right. \right)\, \mathrm{d}y
		\label{Eq:f_Y_1_L_2_s1}
	\end{align}
	\hrulefill
\end{figure*}
By employing~\cite[eq. (9.31/1)]{B:Gra_Ryz_Book}, we can write~\eqref{Eq:f_Y_1_L_2_s1} as in~\eqref{Eq:f_Y_1_L_2_s2}, given at the top of the next page. 
\begin{figure*}
	\begin{align}
		f_{Y_1}(x;L=2) = \frac{\prod_{i=1}^{2}\xi_i}{\prod_{i=1}^2\Gamma\left(\alpha_i\right) \Gamma\left(\beta_i\right)} x^{-1}
		\int_{0}^{\infty} y^{-1} \mathrm{G}_{1,3}^{3, 0}\left(\frac{\alpha_1\beta_1 y}{A_{o,1} \Omega_1}\left|\begin{array}{c} \xi_1+1 \\ \alpha_1, \beta_1, \xi_1 \end{array}\right. \right) 
		\mathrm{G}_{3,1}^{0, 3}\left(\frac{A_{o,2} \Omega_2 y}{\alpha_2\beta_2 x}\left|\begin{array}{c} 1-\alpha_2, 1-\beta_2, 1-\xi_2  \\ 1-\xi_2  \end{array}\right. \right)\, \mathrm{d}y
		\label{Eq:f_Y_1_L_2_s2}
	\end{align} 
	\hrulefill
\end{figure*}
Finally, by applying~\cite[eq. (2.24.5/3)]{B:Prudnikov_v3} in~\eqref{Eq:f_Y_1_L_2_s2}, we~get
\begin{align}
	f_{Y_1}(x;L=2) &= \frac{\prod_{i=1}^{2}\xi_i}{\prod_{i=1}^2\Gamma\left(\alpha_i\right) \Gamma\left(\beta_i\right)} x^{-1} 
	\nonumber \\ & \hspace{-1.8cm}\times
	\mathrm{G}_{2, 6}^{6, 0}\left(\frac{\prod_{i=1}^{2}\alpha_i\beta_i}{A_{o,1}\prod_{i=1}^{2}\Omega_i}y\left| \begin{array}{c} \xi_1+1, \xi_2+1 \\ \alpha_1, \beta_1, \xi_1, \alpha_2, \beta_2, \xi_2 \end{array}\right. \right).
	\label{Eq:f_Y_1_L_2_s3}
\end{align} 	
By recurrently conducting this procedure, for $L=3, 4, \cdots$, we prove that $f_{Y_1}(x)$ can be obtained as in~\eqref{Eq:f_Y_1}, given at the top of the next page.
\begin{figure*}
	\begin{align}
		f_{Y_1}(x) &= \frac{\prod_{i=1}^{L}\xi_i}{\prod_{i=1}^L\Gamma\left(\alpha_i\right) \Gamma\left(\beta_i\right)} x^{-1} 
		\mathrm{G}_{L, 3L}^{3L, 0}\left(\frac{\prod_{i=1}^{L}\alpha_i\beta_i}{\prod_{i=1}^{L}\Omega_i A_{o,i}}x\left| \begin{array}{c} \xi_1+1, \xi_2+1, \cdots, \xi_L+1 \\ \alpha_1, \cdots, \alpha_L, \beta_1, \cdots, \alpha_L, \xi_1, \xi_2, \cdots, \xi_L \end{array}\right. \right).
		\label{Eq:f_Y_1}
	\end{align} 
	\hrulefill
\end{figure*}
Moreover, by following similar steps, and employing~\cite[ch. 2.6]{Mathai1973} as well as~\cite[eq. (2.24.5/3)]{B:Prudnikov_v3}, the PDF of $Y_2$ can be evaluated~as in~\eqref{Eq:Z_1}, given at the top of the next page. 
\begin{figure*}
\begin{align}
	f_{Y_{2}}(x) = \frac{\mathrm{G}_{0, 2 \left(N-L\right)}^{2\left(N-L\right), 0}\left[ x \prod_{i=L+1}^{N} \frac{\alpha_i \beta_i}{\Omega_i} \Big| \alpha_{L+1}, \beta_{L+1}, \alpha_{L+2}, \beta_{L+2} \cdots, \alpha_N, \beta_N \big. \right]}{x\prod_{i=L+1}^{N} \Gamma\left(\alpha_i\right) \Gamma\left(\beta_i\right)},
	\label{Eq:Z_1} 
\end{align}  
\hrulefill
\end{figure*}

Note that $Y_1$ and $Y_2$ are independent RVs; thus, the PDF of $Z=Y_1\,Y_2$ can be evaluated~as 
\begin{align}
	f_{Z}(x) = \int_{0}^{\infty} \frac{1}{y} f_{Y_{1}}(y) \, f_{Y_{2}}\left(\frac{x}{y}\right)\, \mathrm{d}y,
\end{align}
which by substituting~\eqref{Eq:f_Y_1} and~\eqref{Eq:Z_1}, can be written as in~\eqref{Eq:f_Z_s1}, given at the top of the next page.
\begin{figure*}
	\begin{align}
		f_{Z}(x) = \frac{\prod_{i=1}^{L}\xi_i}{\prod_{i=1}^N\Gamma\left(\alpha_i\right) \Gamma\left(\beta_i\right)} 
		x^{-1}
		&\int_{0}^{\infty} y^{-1} 
		\mathrm{G}_{L, 3L}^{3L, 0}\left(\frac{\prod_{i=1}^{L}\alpha_i\beta_i}{\prod_{i=1}^{L}\Omega_i A_{o,i}}y\left| \begin{array}{c} \xi_1+1, \xi_2+1, \cdots, \xi_L+1 \\ \alpha_1, \cdots, \alpha_L, \beta_1, \cdots, \alpha_L, \xi_1, \xi_2, \cdots, \xi_L \end{array}\right. \right)
		\nonumber \\ & \times
		\mathrm{G}_{0, 2\left(N-L\right)}^{2\left(N-L\right), 0}\left( \frac{x}{y} \prod_{i=L+1}^{N} \frac{\alpha_i \beta_i}{\Omega_i} \Big| \alpha_{L+1}, \beta_{L+1}, \alpha_{L+2}, \beta_{L+2} \cdots, \alpha_{N}, \beta_{N} \big. \right) \,\mathrm{d}y
		\label{Eq:f_Z_s1}
	\end{align}
	\hrulefill
\end{figure*}
By employing~\cite{WS:mathematica_function},~\eqref{Eq:f_Z_s1} can be rewritten in a closed-form as in~\eqref{Eq:f_Z_s2}, given at the top of the next page.
\begin{figure*}
	\begin{align}
		f_{Z}(x) &= \frac{\xi^L}{\prod_{i=1}^N\Gamma\left(\alpha_i\right) \Gamma\left(\beta_i\right)} 
		x^{-1}
		\int_{0}^{\infty} y^{-1} 
		\mathrm{G}_{L, 3L}^{3L, 0}\left(\frac{\prod_{i=1}^{L}\alpha_i\beta_i}{\prod_{i=1}^{L}\Omega_i A_{o,i}}y\left| \begin{array}{c} \xi_1+1, \xi_2+1, \cdots, \xi_L+1 \\ \alpha_1, \cdots, \alpha_L, \beta_1, \cdots, \alpha_L, \xi_1, \xi_2, \cdots, \xi_L \end{array}\right. \right)
		\nonumber \\ & \times
		\mathrm{G}_{2\left(N-L\right),0}^{0,2\left(N-L\right),}\left( \frac{y}{x} \prod_{i=L+1}^{N} \frac{\Omega_i}{\alpha_i \beta_i} \Big| 1-\alpha_{L+1}, 1-\beta_{L+1}, 1-\alpha_{L+2}, 1-\beta_{L+2} \cdots, 1-\alpha_{N}, 1-\beta_{N} \big. \right) \,\mathrm{d}y
		\label{Eq:f_Z_s2}
	\end{align}
	\hrulefill
\end{figure*}
Finally, by employing~\cite[eq. (2.24.5/3)]{B:Prudnikov_v3}, we~get~\eqref{Eq:f_Z_final}, given at the top of the next page. 
\begin{figure*}
	\begin{align}
		f_{Z}(x) &= 
		\frac{\prod_{i=1}^{L}\xi_i}{\prod_{i=1}^N\Gamma\left(\alpha_i\right) \Gamma\left(\beta_i\right)} 
		x^{-1}
		\mathrm{G}_{ L,2N+L}^{ 2N+L,0}\left( \frac{1}{\prod_{i=1}^{N}\frac{\Omega_i}{\alpha_i\beta_i}\prod_{i=1}^{L}A_{o,i}} x
		\left| 
		\begin{array}{c}
			1+\xi_1, 1+\xi_2, \cdots, 1+\xi_L \\
			\alpha_{1},\cdots,\alpha_{N}, \beta_{1}, \cdots, \beta_{N}, \xi_1, \cdots, \xi_L
		\end{array}\right. \right)
		\label{Eq:f_Z_final}
	\end{align}
	\hrulefill
\end{figure*}
The CDF of $Z$ can be evaluated~as
\begin{align}
	F_{Z}(x) = \int_{0}^{x} f_{Z}(y)\,\mathrm{d}y,
\end{align} 
which, by applying~\cite[eq. (2.24.5/3)]{B:Prudnikov_v3}, returns~\eqref{Eq:F_Z_final}, given at the top of the next page.
\begin{figure*}
	\begin{align}
		F_{Z}(x) = \frac{\prod_{i=1}^{L}\xi_i}{\prod_{i=1}^N\Gamma\left(\alpha_i\right) \Gamma\left(\beta_i\right)} \mathrm{G}_{L, 1}^{2N+L, L}\left(\frac{1}{\prod_{i=1}^{N}\frac{\Omega_i}{\alpha_i\beta_i}\prod_{i=1}^{L}A_{o,i}} x
		\left|\begin{array}{c}1, \xi_1+1, \cdots, \xi_L+1 \\  \alpha_1, \cdots, \alpha_N, \beta_1, \cdots, \beta_N, \xi_1, \cdots, \xi_L, 0  \end{array}\right.\right)
		\label{Eq:F_Z_final}
	\end{align}
	\hrulefill
\end{figure*}
From~\eqref{Eq:F_Z_final}, we can straightforwardly express the CDF of $A_t$~as
\begin{align}
	F_{A_{t}}(x) = F_{Z}\left(\frac{x}{\prod_{i=1}^{N}g_i}\right),
\end{align} 
which leads to~\eqref{Eq:F_At_final}, given at the top of the next page. 
\begin{figure*}
	\begin{align}
		F_{A_{t}}(x) = \frac{\prod_{i=1}^{L}\xi_i}{\prod_{i=1}^N\Gamma\left(\alpha_i\right) \Gamma\left(\beta_i\right)} \mathrm{G}_{L, 1}^{2N+L, L}\left(\frac{1}{\prod_{i=1}^{N}\frac{\Omega_i}{\alpha_i\beta_i}\prod_{i=1}^{L}A_{o,i}} \frac{x}{\prod_{i=1}^{N}g_i}
		\left|\begin{array}{c}1, \xi_1+1, \cdots, \xi_L+1 \\  \alpha_1, \cdots, \alpha_N, \beta_1, \cdots, \beta_N, \xi_1, \cdots, \xi_L, 0  \end{array}\right.\right)
		\label{Eq:F_At_final}
	\end{align}
	\hrulefill
\end{figure*}
Finally, by applying~\eqref{Eq:A_t} in~\eqref{Eq:P_o_THz_s2}, we get~\eqref{Eq:P_o_THz_s3}. This concludes the proof.

\balance
\bibliographystyle{IEEEtran}
\bibliography{IEEEabrv,References}

\end{document}